\documentclass[twocolumn,trackchanges]{aastex63}

\newcommand{\Rmnum}[1]{\expandafter\@slowromancap\romannumeral #1@}
\usepackage{graphicx,epsfig,fancyhdr,psfig,rotating,amsmath,epsf,txfonts,natbib,epstopdf,multirow}
\def \kms {{\rm km\;s$^{-1}$}}

\def \arcsec {$^{''}$}
\def \siiv {Si\,{\sc iv}}

\def \oiv {O\,{\sc iv}}

\def\paper1{{ Paper I}}
\graphicspath{{./figs/}} 
\def \halpha {{H$\alpha$}}
\received{2020}
\revised{ 2020}
\accepted{2020}
\begin{document}
\title{Heating at the remote footpoints as a brake on jet flows along loops in the solar atmosphere}
\correspondingauthor{Zhenghua Huang}
\email{z.huang@sdu.edu.cn}

\author{Zhenghua Huang}
\affiliation{Shandong Provincial Key Laboratory of Optical Astronomy and Solar-Terrestrial Environment, Institute of Space Sciences, Shandong University, Weihai, 264209 Shandong, China}

\author{Qingmin Zhang}
\affiliation{Key Laboratory of Dark Matter and Space Astronomy, Purple Mountain Observatory, CAS, Nanjing 210033, China}

\author{Lidong Xia}
\affiliation{Shandong Provincial Key Laboratory of Optical Astronomy and Solar-Terrestrial Environment, Institute of Space Sciences, Shandong University, Weihai, 264209 Shandong, China}

\author{Bo Li}
\affiliation{Shandong Provincial Key Laboratory of Optical Astronomy and Solar-Terrestrial Environment, Institute of Space Sciences, Shandong University, Weihai, 264209 Shandong, China}

\author{Zhao Wu}
\affiliation{Shandong Provincial Key Laboratory of Optical Astronomy and Solar-Terrestrial Environment, Institute of Space Sciences, Shandong University, Weihai, 264209 Shandong, China}

\author{Hui Fu}
\affiliation{Shandong Provincial Key Laboratory of Optical Astronomy and Solar-Terrestrial Environment, Institute of Space Sciences, Shandong University, Weihai, 264209 Shandong, China}

%\author{
%\sc{Zhenghua Huang\altaffilmark{1}, Qing-Min Zhang\altaffilmark{2}, Lidong Xia\altaffilmark{1}, Bo Li\altaffilmark{2}, Hui Fu\altaffilmark{1}}
%}
%\altaffiliation{1}{Shandong Provincial Key Laboratory of Optical Astronomy and Solar-Terrestrial Environment, Institute of Space Sciences, Shandong University, Weihai, 264209 Shandong, China; {\it z.huang@sdu.edu.cn}}
%\altaffiliation{2}{Key Laboratory for Dark Matter and Space Science, Purple Mountain Observatory, CAS, Nanjing 210034, People's Republic of China}

%\author{
%Zhenghua Huang, Qing-Min Zhang, Lidong Xia, Bo Li, Hui Fu
%}
%\affiliation{Shandong Provincial Key Laboratory of Optical Astronomy and Solar-Terrestrial Environment, Institute of Space Sciences,
%Shandong University, Weihai, 264209 Shandong, China; {\it z.huang@sdu.edu.cn}
%Key Laboratory for Dark Matter and Space Science, Purple Mountain Observatory, CAS, Nanjing 210034, People's Republic of China
%}
%$\lesssim$

\begin{abstract}
We report on observations of a solar jet propagating along coronal loops taken by the Solar Dynamics Observatory (SDO), the Interface Region Imaging Spectragraph (IRIS) and 1-m New Vacuum Solar Telescope (NVST).
The ejecta of the jet consist of multi-thermal components and propagate with a speed greater than 100\,\kms.
Brightenings are found in the remote footpoints of the coronal loops having compact and round-shape in the \halpha\ images.
The emission peak of the remote brightening in the Atmospheric Imaging Assembly (AIA) 94\,\AA\ passband lags 60\,s behind that in the jet base.
The brightenings in the remote footpoints are believed to be consequences of heating by nonthermal electrons, MHD waves and/or conduction front generated by the magnetic reconnection processes of the jet.
The heating in the remote footpoints leads to extension of the brightening along the loops toward the jet base, which is believed to be the chromospheric evaporation.
This apparently acts as a brake on the ejecta, leading to a deceleration in the range from 1.5 to 3\,km\,s$^{-2}$ with an error of $\sim1.0$\,km\,s$^{-2}$ when the chromospheric evaporation and the ejecta meet at locations near the loop apexes.
The dynamics of this jet allows a unique opportunity to diagnose the chromospheric evaporation from the remote footpoints, from which we deduce a velocity in the range of 330--880\,\kms.
\end{abstract}

%\maketitle

\section{Introduction}
\label{sect_intro}
Solar jets are ejections of solar plasma in a manner that are forced out of a small opening compared to their lengths. 
They are ubiquitous in the solar atmosphere, such as chromosphere spicules\,\citep[e.g.][etc.]{2000SoPh..196...79S,2007PASJ...59S.655D,2012ApJ...750...16Z,2012ApJ...759...18P,2019Sci...366..890S}, transition region network jets\,\citep[e.g.][etc.]{2014Sci...346A.315T,2016SoPh..291.1129N,2019SoPh..294...92Q} and coronal jets\,\citep[see a review by][]{2016SSRv..201....1R}.
These jets are not only one of the basic components of the solar atmosphere, but also play a crucial role in mass and energy transformation and transportation in the solar atmosphere.
For example, spicules and network jets are believed to play an important role in heating the corona and balancing the mass loss by solar wind\,\citep[see e.g.][]{2011Sci...331...55D,2014Sci...346A.315T}.

\par
In particular, coronal jets have been studied intensively since they were first identified in the early 1990s by the soft X-ray Telescope onboard \textit{Yohkoh}\,\citep{1992PASJ...44L.173S}.
In \textit{Yohkoh} soft X-ray observations, coronal jets normally have lengths ranging from a few $10^4$\,km to $4\times10^5$\,km, widths from $5\times10^3$\,km to $10^5$\,km and lifetimes from a few minutes to a few hours\,\citep{1996PASJ...48..123S}.
The kinetic energy of the coronal jets is estimated to be $10^{25}$--$10^{28}$\,ergs\,\citep{1992PASJ...44L.173S}.
The observations taken by subsequent missions, such as \textit{SOHO}, \textit{TRACE}, \textit{Hinode}, \textit{STEREO} and \textit{SDO}, have revealed that coronal jets can also be seen in the EUV passbands and they contain both hot and cool components\,\citep[see][and the references therein]{2016SSRv..201....1R}.

\par
Coronal jets are believed to be consequences of magnetic reconnection\,\citep{1995Natur.375...42Y,2017Natur.544..452W}.
Many studies have focused on observations of the evidence and the dynamic processes of magnetic reconnection in coronal jets.
Many coronal jets tend to have an inverted-Y shape\,\citep[see e.g.][]{1992PASJ...44L.173S,2011A&A...526A..19M}, suggesting magnetic reconnection between small loops and (quasi-) open field lines\,\citep{1995Natur.375...42Y}.
Moreover, magnetic cancellations, which are one of the signatures of magnetic reconnection\,\citep[e.g.][]{2018ApJ...862L..24P,2019ApJ...872...32S,2020ApJ...891...52S}, have also been found to be closely associated with the occurrences of coronal jets\,\citep[see e.g.][etc.]{2012A&A...548A..62H,2017ApJ...844..131P,2019ApJ...882...16M}.
In some coronal jets, magnetic reconnections are also evidenced by the presence of blobs in current sheets\,\citep{2014A&A...567A..11Z,2016SoPh..291..859Z,2019ApJ...870..113Z}.
Recent studies reveal that the magnetic reconnection in coronal jets is triggered by mini-filament eruptions\,\citep[e.g.][etc.]{2015Natur.523..437S,2016ApJ...821..100S,2016ApJ...830...60H,2018ApJ...854...80H,2018ApJ...859....3M,2019ApJ...871..220S,Hong_2019,2019ApJ...883..104S,2019ApJ...887..220Y}.
This suggests that coronal jets involve physical processes that couple the solar atmosphere from the chromosphere to the corona.

\par
In most coronal jets, part of the ejecta tend to fall back toward the solar surface.
In these cases, the plasma flows may have a parabolic trajectory\,\citep[see examples in][though the falling back motions are not their topics]{2019ApJ...887..154L}.
Here, we report on observations of a solar jet flowing along closed loops, in which the deceleration of the ejecta was associated with heating in the remote footpoints of the loops.
In what follows, we give the description of the data in Section\,\ref{sect:obs}, the data analysis and results in Section\,\ref{sect:res} and the discussion and conclusions in Section\,\ref{sect:concl}.

\begin{figure*}
\includegraphics[clip,trim=1cm 1.3cm 0cm 0cm,width=\textwidth]{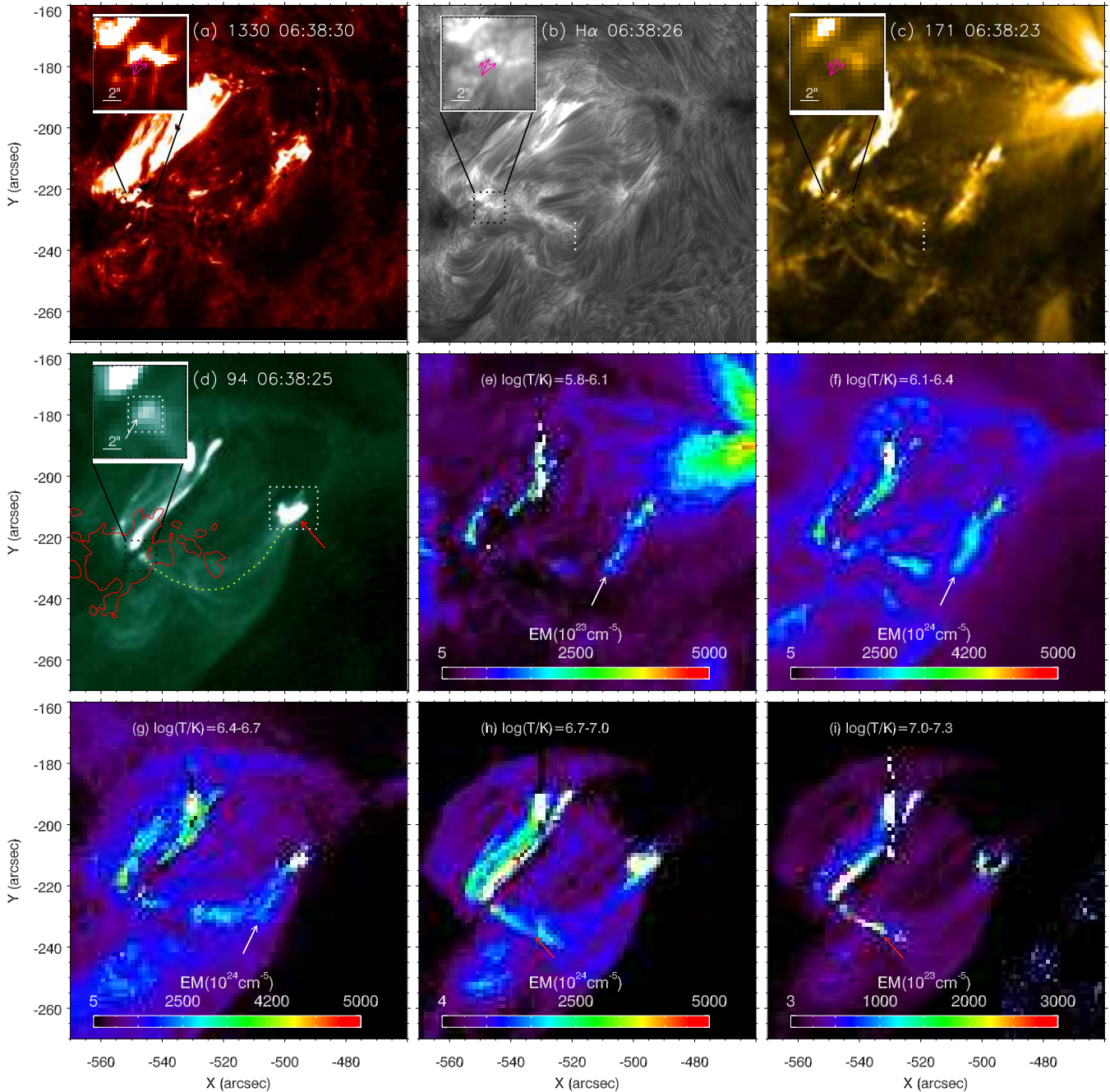}
\caption{The region with jet activity viewed in IRIS SJ 1330\,\AA\ (a), NVST \halpha\ (b), AIA 171\,\AA\ (c), AIA 94\,\AA\ (d) passbands and emission measures derived from the AIA data (e--i).
The jet base is denoted by the arrow in red in panel d.
The remote footpoints of the jet loops (the region enclosed by black dotted lines) are zoomed in and displayed in the sub-images on panels (a)--(d).
The arrows in purple in the sub-images of panels a--c point to locations of three compact bright kernels seen in the \halpha\ image.
The arrow in white in the sub-image of panel d points to the bright core of the remote footpoints in the AIA 94\,\AA\ images.
The contours in red in panel d mark the places around the remote footpoints where the HMI line-of-sight magnetic flux density is $-500$\,Mx\,cm$^{-2}$ (also see Figure\,\ref{fig:mag}).
The white dotted lines in panel d enclose the region of jet base and the remote footpoints, from which the light curves are shown in Figure\,\ref{fig:lc}.
The dotted line in yellow in panel (d) marks the path of a set of jet loops, from which the time-distance plots in Figure\,\ref{fig:stva} are obtained.
The white dotted lines in panel (b--c) indicate the locations from which the IRIS spectra shown in Figure\,\ref{fig:spden} are taken.
The white arrows in panels (e--g) point to the jet plasma and the red arrows in panels (h--i) point to the bright features extending from the remote footpoint.
An animation of this figure are provided online.
}
\label{fig:mulimgs}
\end{figure*}

\begin{figure}
\includegraphics[clip,trim=0cm 1cm -0.4cm 0cm,width=\linewidth]{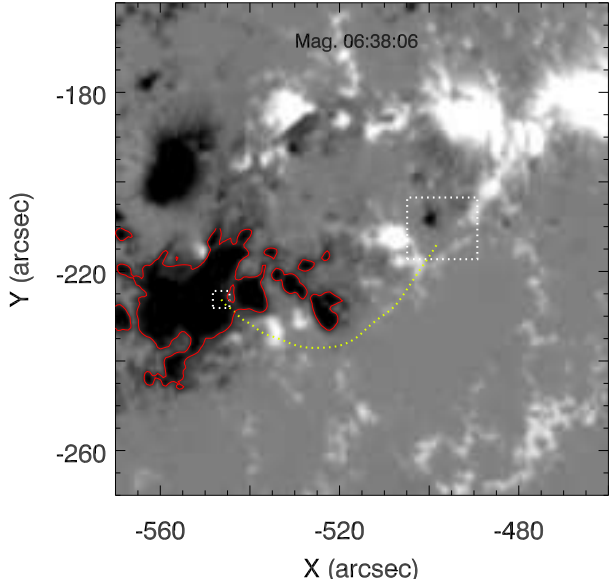}
\caption{HMI line-of-sight magnetogram of the region shown in Figure\,\ref{fig:mulimgs}.
The image is artificially saturated at $-800$\,Mx\,cm$^{-2}$ (black) and 800\,Mx\,cm$^{-2}$ (white).
The contours in red mark the places around the remote footpoints where the HMI line-of-sight magnetic flux density is $-500$\,Mx\,cm$^{-2}$.
The white dotted lines enclose the jet base region and remote footpoint region as shown in Figure\,\ref{fig:mulimgs}d.
The yellow dotted line outlines the path of the jet loops as shown in Figure\,\ref{fig:mulimgs}d.
}
\label{fig:mag}
\end{figure}

\begin{figure}
\includegraphics[clip,trim=0cm 0.5cm 0cm 0cm,width=\linewidth]{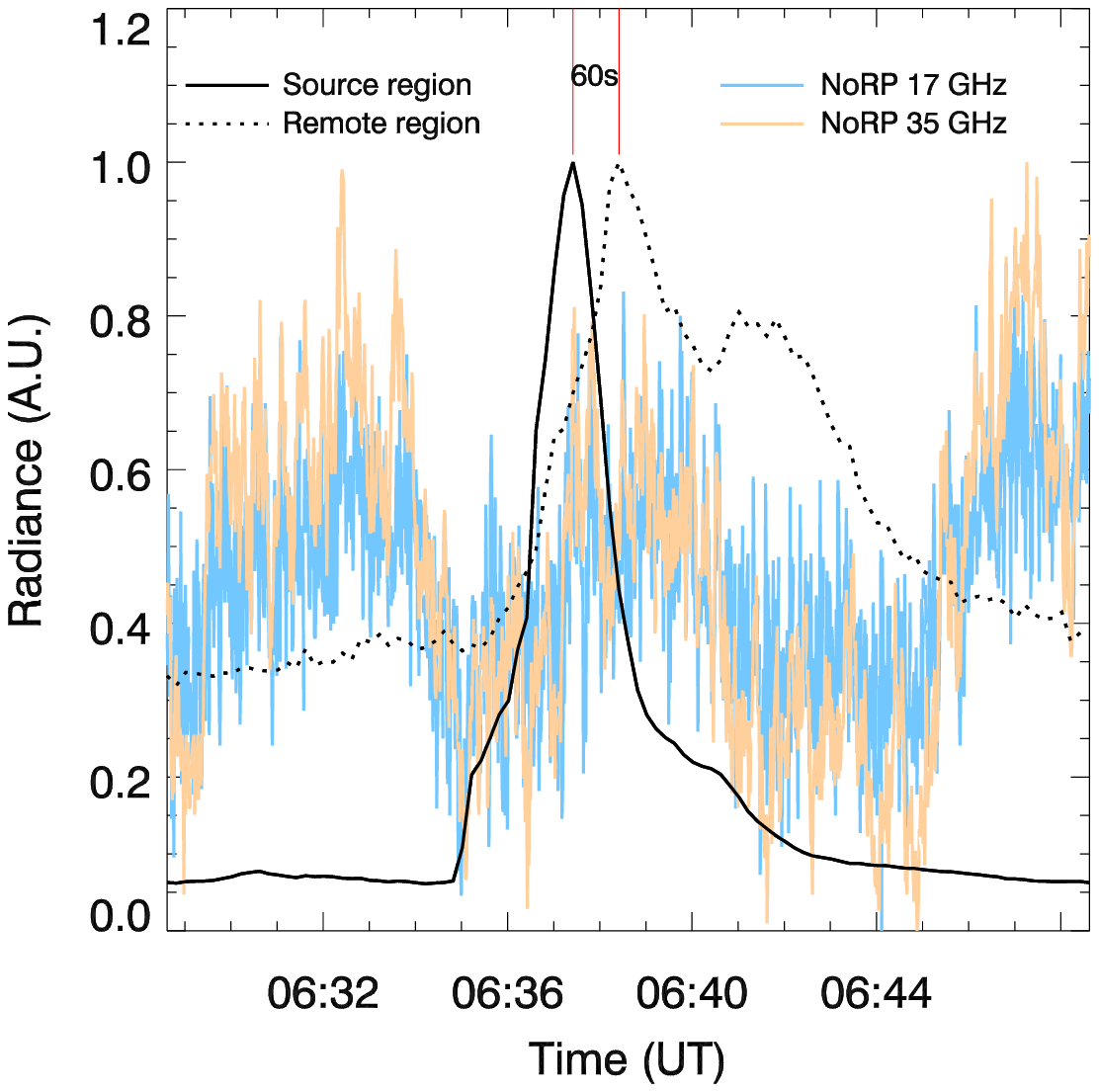}
\caption{The light curves of the emission in the AIA 94\,\AA\ passband of the jet base (solid line) and remote footpoints (dotted line) as denoted in Figure\,\ref{fig:mulimgs}d and the relative variations of NoRP microwave flux at 17 GHz (blue) and 35 GHz (orange).
The two red lines mark the locations of the peaks of the two AIA 94\,\AA\ light curves, which have a time difference of 60\,s.
}
\label{fig:lc}
\end{figure}

\section{Data description} 
\label{sect:obs}
The observations analysed here were taken by the Atmospheric Imaging Assembly\,\citep[AIA,][]{2012SoPh..275...17L} and the Helioseismic and Magnetic Imager\,\citep[HMI,][]{2012SoPh..275..207S} aboard the Solar Dynamics Observatory\,\citep[SDO,][]{2012SoPh..275....3P}, the Interface Region Imaging Spectragraph\,\citep[IRIS,][]{2014SoPh..tmp...25D} and the 1-m New Vacuum Solar Telescope\,\citep[NVST,][]{2014RAA....14..705L} of the Fuxian Solar Observatory (FSO) on 4 May 2015. 

\par
The AIA data used in this study include images taken in the EUV passbands around 94\,\AA, 131\,\AA, 171\,\AA, 193\,\AA, 211\,\AA\ and 335\,\AA.
The temperature response functions of these passbands peaks at $6.3\times10^6$\,K, $1.0\times10^7$\,K, $6.3\times10^5$\,K, $1.3\times10^6$\,K, $2.0\times10^6$\,K and $2.5\times10^6$\,K, respectively \,\citep{2012SoPh..275...17L}
The cadences of the images of both the passbands are 12\,s and the pixel sizes are 0.6\,\arcsec$\times$0.6\arcsec.
The line-of-sight magnetograms taken by HMI are used, which have a cadence of 45\,s and a pixel size of 0.6\,\arcsec$\times$0.6\arcsec.
Both AIA and HMI data have been reduced by the standard procedure of \textit{aia\_prep.pro} in the \textit{solarsoft}.

\par
The IRIS spectrograph was running in a ``large coarse 8-step raster'' mode with a 2\arcsec\ step size and 8\,s exposure time.
It obtained 80 rasters of the spectral data in total and the cadence between each raster is 72\,s.
The spectrograph slit has 0.35\arcsec\ width and the pixel size along the slit axis is 0.33\arcsec because the data have been binned by a factor of 2 aboard before being transferred to the ground.
The IRIS slit-jaw (SJ) imager took observations in the passbands of 1330\,\AA\ and 2796\,\AA\ with a cadence of 18\,s and a pixel size of 0.17\arcsec$\times$0.17\arcsec.
We use the IRIS level 2 data that have been calibrated by the operation team and are suitable for science analyses.
To do the radiometric calibration, we follow the procedures given in the IRIS technical note 24.

\par
The analysed data also include observations from the NVST, which were taken in \halpha\ line centre with a bandpass of 0.25\,\AA.
The \halpha\ data have a pixel size of 0.17\arcsec$\times$0.17\arcsec.
The data have been reduced and reconstructed using Speckle technics\,\citep[see details in][]{2016NewA...49....8X}.
The reconstructed data have a cadence of about 12\,s.

\par
The data taken from different instruments are aligned by using the images of passbands having close representative temperatures.
For this purpose, the images taken with the AIA 1600\,\AA\ and 304\,\AA\ passbands have also been used.
Moreover, the calibrated full-disc microwave flux (intensity) at 17\,GHz and 35\,GHz taken by the Nobeyama Radio Polarimeters\,\citep[NoRP,][]{1985PASJ...37..163N} are also used to determine the variations of the nonthermal radiations.
The NoRP data have a cadence of 1\,s.

\begin{figure*}
\includegraphics[clip,trim=0cm 0cm 0cm 0cm,width=\textwidth]{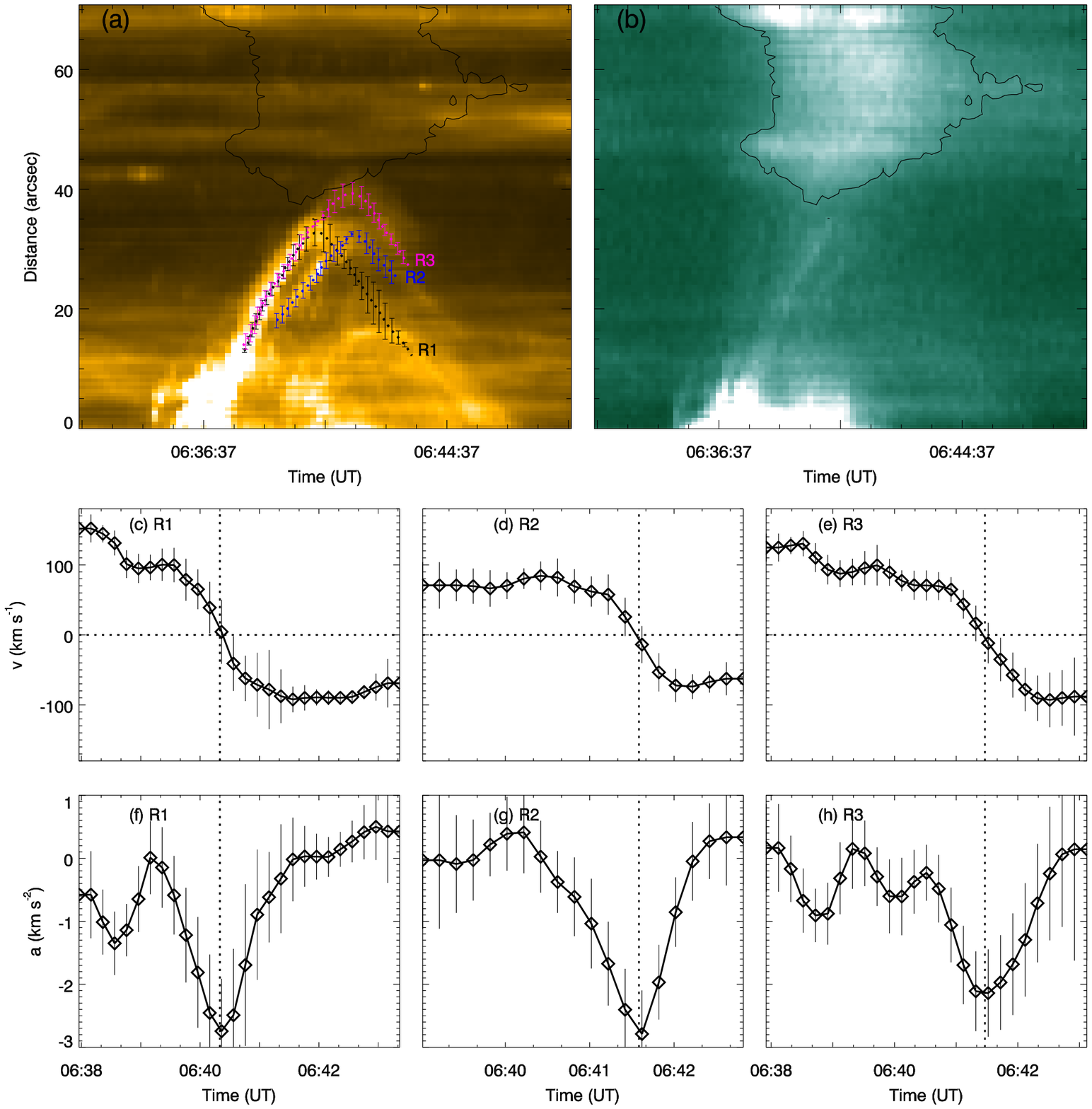}
\caption{Analyses of the velocity and acceleration of the jet along the jet loops marked by yellow dotted line in Figure\,\ref{fig:mulimgs}d.
Panel (a): time-distance plot based on the AIA 171\,\AA\ observations.
Three trajectories of the ejecta are marked by black (``R1''), blue (``R2'') and purple (``R3'') dotted lines.
The error bars of the trajectories are determined from the spatial ranges in which the intensity drops to 90\% of the center.
The contours (black solid lines) mark the extension of the brightness from the remote footpoints based on the AIA 94\,\AA\ observations.
Panel (b): time-distance plot based on the AIA 94\,\AA\ observations.
The contours (black solid lines) are the same as that in panel (a).
Panels (c)--(e): variations of the velocities of the ejecta along the trajectories of ``R1''--``R3'' based on the AIA 171\,\AA\ time-distance plot shown in panel (a).
The dotted lines mark the locations of zero velocity.
The error bars are 1$\sigma$ uncertainty of the linear fits based on the trajectories and the associated errors.
Panels (f)--(h): variations of accelerations of the ejecta based on the velocity variations shown in panels (c)--(e).
The dotted lines mark the locations of zero velocity.
The error bars are 1$\sigma$ uncertainty of the linear fits based on the velocity curves and the associated errors.
}
\label{fig:stva}
\end{figure*}

\section{Data analyses and results}
\label{sect:res}
We first derive the emission measures (EM) of the jet based on the observations of the AIA 94\,\AA, 131\,\AA, 171\,\AA, 193\,\AA, 211\,\AA\ and 335\,\AA\ passbands with the differential emission measure (DEM) code developed and optimized by \citet{2015ApJ...807..143C} and \citet{2018ApJ...856L..17S}.
To reduce noise while constructing an EM image of the region, we have binned the AIA data by 2\,pixels$\times$2\,pixels, which leaves a spatial sampling of 1.2\arcsec$\times$1.2\arcsec\ for the constructed EM images. 
The evolution of the region where the jet took place as seen in IRIS SJ 1330\,\AA, the NVST \halpha, the AIA 171\,\AA\ and the AIA 94\,\AA\ images and the constructed emission measures at a few ranges of temperatures are shown in Figure\,\ref{fig:mulimgs} and the associated animation.

\par
The solar jet initiates at around 06:35\,UT (see its base region denoted by the red arrow in Figure\,\ref{fig:mulimgs}d).
The system seems to be initially destabilised by a  GOES B9.7 flare (see its main ribbons as the bright region at coordinates of X=$-560\sim-520$\arcsec\ and Y=$-220\sim-180$\arcsec\ in Figure\,\ref{fig:mulimgs}, and the flare started from 06:31:20\,UT, peaked at 06:37:14\,UT and ended at 06:48:16\,UT as given in the RHESSI flare list\footnote{\url{https://hesperia.gsfc.nasa.gov/rhessi3/data-access/rhessi-data/flare-list/index.html}}).
In the base region of the jet, we can see the eruption of a dark filament-like feature in AIA 171\,\AA\ and \halpha\ images (see the animation associated with Figure\,\ref{fig:mulimgs}b\&c).
In the AIA 94\,\AA\ images, we can see that multiple coronal loops are also rooted in the base of the jet (see Figure\,\ref{fig:mulimgs}d and one example is denoted by yellow dotted line that is about 70\arcsec\ in length).
Hereafter, these loops are termed ``jet loops'', and their footpoints other than the base of the jet are termed ``remote footpoints''.
The HMI magnetogram reveals a magnetic structure with a patch of negative polarity surrounded by positive polarities in the base of the jet (see Figure\,\ref{fig:mag}). 
The coronal loops seen in the AIA 94\,\AA\ channel connect the base of the jet and the negative polarities spread in the regions toward the east (see the red contours shown in Figure\,\ref{fig:mulimgs}d and Figure\,\ref{fig:mag}).

\par
The jet activity begins with appearance of two bright cores in the base of the jet as seen in the IRIS SJ 1330\,\AA\ image at 06:34:53\,UT.
The brightening in the base evolves into a circular ribbon seen in the \halpha\ images (see Figure\,\ref{fig:mulimgs}b) and arcade-like feature in the AIA 94\,\AA\ images (see Figure\,\ref{fig:mulimgs}d).
The increase of the bright region in the base of the jet is then followed by plasma ejections along the coronal loops (see the animation associated with Figure\,\ref{fig:mulimgs}).
Along with the plasma ejections, we can also see that dark mini-filament threads are lifted off in the \halpha\ images, indicating of an eruption of mini-filaments that agrees with the previous studies\,\citep[see e.g.][]{2015Natur.523..437S}.
In the \halpha\ images, we also see that the jet front appears to be brighter than the background.
This indicates that there is a cool counterpart (as seen in \halpha) of the jet.
The multi-thermal nature of the jet is also confirmed by the constructed EMs, which show clear structures with enhanced emission in the temperatures from log(T/K)=5.8 to log(T/K)=6.7 (see the structures denoted by the arrows in Figures\,\ref{fig:mulimgs}e--g).

\par
Almost at the same time when the jet was leaving the bright base, we see brightenings in the remote footpoints of the jet loops (see the sub-image shown in each panel of Figure\,\ref{fig:mulimgs} and the associated animation).
In the \halpha\ images, we can see multiple round-shaped bright kernels with a subarcsecond size, which are isolated from each other (see three clear examples denoted by purple arrows in Figure\,\ref{fig:mulimgs}b).
The flaring time of each bright kernel appears to be different from the others.
In the IRIS 1330\,\AA\ images, bright features with similar structure are also occurring at the same location (see Figure\,\ref{fig:mulimgs}a).
However, they are not as isolated as that shown in the \halpha\ images because the saturation in the IRIS observations.
Similar bright features can also be seen in the AIA 171\,\AA\ images (see Figure\,\ref{fig:mulimgs}c).
The morphology of the bright feature in the AIA 94\,\AA\ images is a contrast to that shown in the others diplayed in Figure\,\ref{fig:mulimgs}.
The bright features in the AIA 94\,\AA\ images appears to have only one bright kernel that is located in the place surrounded by those seen in the other three passbands (see Figure\,\ref{fig:mulimgs}d).
We further compare the light-curves of the AIA 94\,\AA\ emissions at the jet base and the remote footpoints (Figure\,\ref{fig:lc}). 
There is a time lag of 60\,s between the peak in the emission of the remote footpoints and that of the jet base (see Figure\,\ref{fig:lc}).
Moreover, in the AIA 94\,\AA\ images, we can see that the dynamics in the remote footpoints is followed by sudden extension of brightness along the jet loop toward the jet base (see the animation associated with Figure\,\ref{fig:mulimgs}).
The bright feature shows response only in the AIA\,94\,\AA\ and 131\,\AA\ passbands, suggesting its high temperature nature.
The DEM analyses reveal enhanced emissions in the temperatures from log(T/K)=6.7 to log(T/K)=7.3 (see the structures denoted by the arrows in Figures\,\ref{fig:mulimgs}h--i).
These EM images indicate that the feature could be heated to a temperature as high as $10^7$\,K.

\par
The variations of the NoRP microwave flux show clear increase in both frequencies during the time when the remote footpoints are getting brighter in AIA 94\,\AA\ (see the peaks of the NoRP curves around 06:38\,UT in Figure\,\ref{fig:lc}).
This indicates the enhanced nonthermal radiations of the Sun due to nonthermal particles produced in magnetic reconnections\,\citep[e.g.][]{2012JGRA..117.7223W,2015ApJ...806..167G}.
In order to determine whether the enhancement of nonthermal radiation is from the studied event,
we carefully check the AIA full-disc observations between 06:20\,UT and 06:50\,UT using Helioviewer\footnote{\url{https://helioviewer.org}} and the RHESSI flare list.
We found that the peaks of NoRP curves around 06:32\,UT are corresponding to the impulsive phase of the  B9.7 flare mentioned previously and the peaks around 06:47\,UT are associated with another weak flare occurring in a different active region.
During the period between these two small flares, we do not see any other obvious activities on the near side of the Sun, neither in the AIA EUV images nor in the RHESSI records.
Therefore, we believe that the physical processes associated with the jet are responsible for the enhancement in the NoRP flux around 06:38\,UT.

\par
%\subsection{Dynamics of the plasma ejecta}
After 06:35:37\,UT, we observe that the plasma ejecta of the jet are moving along the jet loops toward the remote footpoints (hereafter, we name the motions toward the remote footpoints as upward motion).
It starts to fall back toward the jet base around 06:40\,UT (hereafter, we name the motions toward the jet base as downward motions).
To investigate the dynamics of the plasma ejecta, we produce time-distance plots along the trajectory of the jet loops (see the yellow dotted line in Figure\,\ref{fig:mulimgs}d) using the AIA 171\,\AA\ observations (Figure\,\ref{fig:stva}a) and the AIA 94\,\AA\ observations (Figure\,\ref{fig:stva}b).
The upward motions of the ejecta have speed more than 130\,\kms, and the downward motions reach a speed as high as 100\,\kms\ (both with $1\sigma$ errors of 10--30\,\kms).
Both the upward and downward motions of the plasma can be clearly seen in the AIA 171\,\AA\ (see the dotted line in Figure\,\ref{fig:stva}a), but in the AIA 94\,\AA\ channel the downward one cannot be detected (see Figure\,\ref{fig:stva}b and also the animation associated with Figure\,\ref{fig:mulimgs}).
We can see that the upward ejecta are much brighter than the downward ones.
These facts indicate that the upward ejecta are multi-thermal with plasma as hot as $6.3\times10^6$\,K, whereas the downward ones are much cooler.

\par
In the AIA 94\,\AA\ channel, we can see that the brightening in the remote footpoints also leads to extension of brightness along the loops (see the region indicated by the contour in Figure\,\ref{fig:stva}b and the animation associated with Figure\,\ref{fig:mulimgs}).
This extension of bright features cannot be seen in the cooler temperatures, suggesting that its temperature can reach $6.3\times10^6$\,K (i.e. representative temperature of the AIA 94\,\AA\ channel).
At the points where the extension of the brightness from the remote footpoints approaches the jet ejecta, we see that the jet ejecta start changing their  directions from upward to downward (see Figure\,\ref{fig:stva}a\&b).
This suggests a connection between the brake on the ejecta and the extension of the brightness from the remote footpoints.
We manually select three trajectories of the ejecta based on the AIA 171\,\AA\ time-distance plots (see ``R1--R3'' in Figure\,\ref{fig:stva}a).
In Figure\,\ref{fig:stva}c--e, we show how the velocities change with time along the three trajectories.
It shows that the velocities rapidly change near the points where the velocities are close to zero (i.e. when the ejecta are approaching the extending brightness from the remote footpoints).
This can also be demonstrated by the accelerations derived from the velocity curves in Figure\,\ref{fig:stva}f--h.
The accelerations have positive value if it has direction toward the remote footpoints.
We can see that the decelerations near the point of zero velocity are in the range of 1.5--3\,km\,s$^{-2}$ with a 1$\sigma$ error of 0.8--1\,km\,s$^{-2}$ (see the points near that marked by the dotted lines in Figure\,\ref{fig:stva}f--h).
These are much larger than the gravitational that is less than 0.3\,km\,s$^{-2}$.
It is clear that the velocities of the ejecta change directions in a very short period of time that the gravity itself cannot explain.
This indicates that there are other forces pushing the ejecta in the direction from the remote footpoints toward the jet base.

\par
\begin{figure*}
\includegraphics[clip,trim=0cm 0cm 0cm 0cm,width=\textwidth]{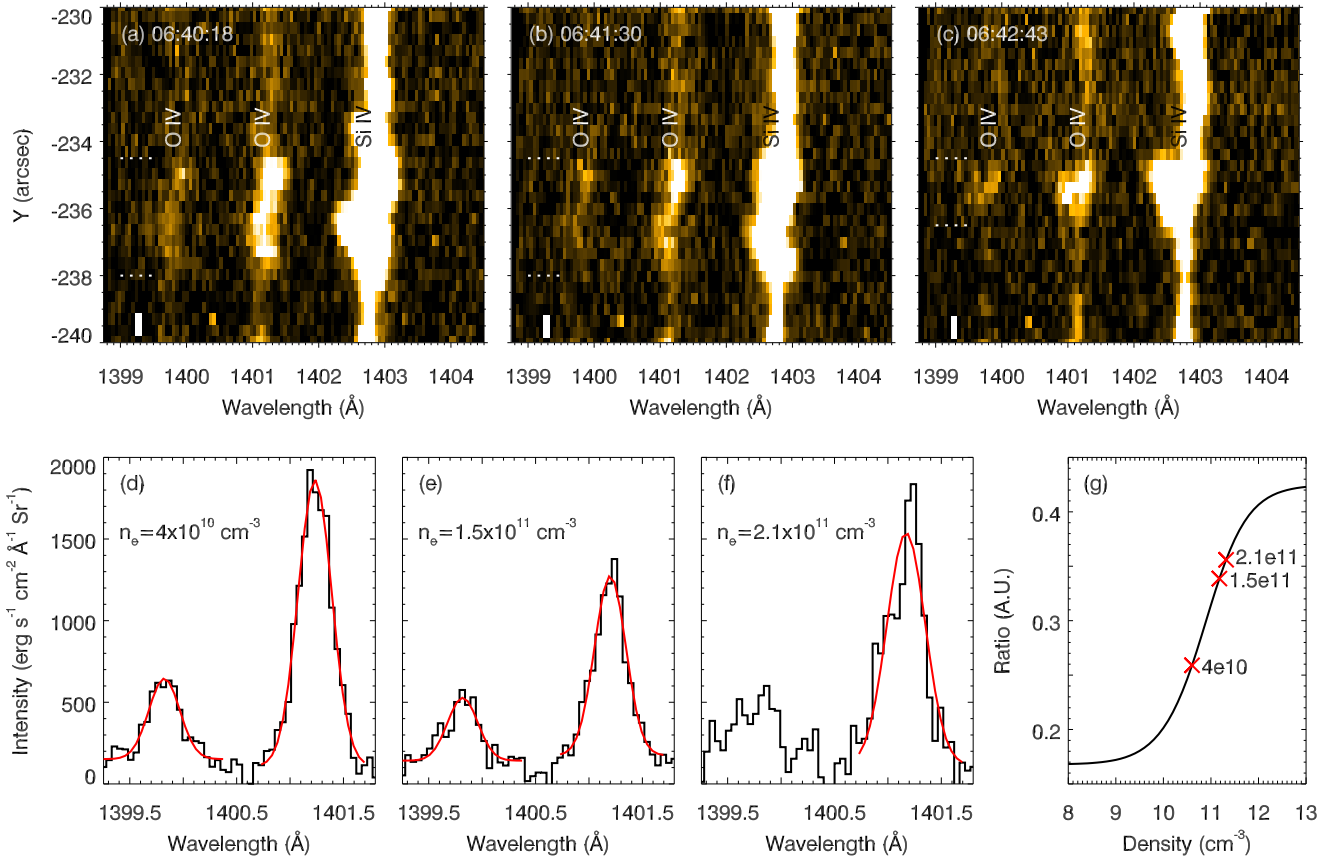}
\caption{Panels (a)--(c): the IRIS slit images in the wavelengths around 1400\,\AA\ of the location marked by dotted lines in Figure\,\ref{fig:mulimgs}b\&c taken at three different time.
Three spectral lines are marked by notes of their emitting ions.
The spatial ranges of the ejecta along the Y axis at the indicated time are marked by the white dotted lines.
Panels (d)--(f): the mean spectral profiles of the ejecta for the times shown in panels (a), (b) and (c), respectively.
The profiles are averaged from the regions marked by the white dotted lines in panels (a)--(c) in the wavelength range of 1399.3--1401.8\,\AA, which includes \oiv\,1399.8\,\AA\ and 1401.2\,\AA.
The black lines are the observations and the red lines are single Gaussian fits of the two \oiv\ lines.
Panel (g): the theoretical profile showing how the line ratio between \oiv\,1399.8\,\AA\ and 1401.2\,\AA\ varies with electron densities.
The line ratios of the three cases shown in panels (d)--(f) are marked by red crosses followed by the corresponding electron densities.
The line ratios of the cases shown in panels (d)\&(e) are obtained through Gaussian fits.
The \oiv\ 1399.8\,\AA\ profile in panel (f) is too weak to have a converged fit, and thus the total intensities in the corresponding wavelength ranges subtracted by the background (derived from the Gaussian fit to the \oiv\,1401.2\,\AA\ profile) are used to calculate the line ratio.
}
\label{fig:spden}
\end{figure*}

The existence of additional forces is also evidenced by the spectral profiles of the ejecta near the loop top.
In Figure\,\ref{fig:spden}a--c, we show the slit images at the wavelengths near 1400\,\AA\ at three points of observing time, which are taken for the places near the loop top (see the white dotted lines in Figure\,\ref{fig:mulimgs}b\&c).
From the animation associated with Figure\,\ref{fig:mulimgs}, we can see the ejecta are in the stage of upward motion at 06:40:18\,UT (Figure\,\ref{fig:spden}a) and 06:41:30\,UT (Figure\,\ref{fig:spden}b) but in the stage of downward motion at 06:42:43\,UT (Figure\,\ref{fig:spden}c).
It clearly shows that the \siiv\ 1403\,\AA\ profiles are blue-shifted in all the three cases.
The profiles at 06:40:18\,UT and 06:42:43\,UT can be blue-shifted to a velocity of 130\,\kms, and that at 06:41:30\,UT is only blue-shifted to less than 100\,\kms.
Since the apparent motions of the ejecta at 06:40:18\,UT and 06:42:43\,UT have opposite directions, the blue-shifted \siiv\ profiles in both cases suggest that the field-of-view of the slit covers a section of the loop at both sides of its apex.
That means the ejecta have passed through the loop apex.
It again indicates existence of forces other than the gravity that push the ejecta toward the jet base.

\par
The ejecta also emit \oiv\ 1399.8\,\AA \& 1401.2\,\AA\ above the instrumental noise level (see Figure\,\ref{fig:spden}d--f).
This allows a diagnostics of the electron densities of the ejecta.
Based on the line ratio vs. electron density curve given by the version 9 of the CHIANTI database\,\citep[][see our Figure\,\ref{fig:spden}g]{1997A&AS..125..149D,Dere_2019}, the electron densities of the ejecta are found to be $4\times10^{10}$\,cm$^{-3}$, $1.5\times10^{11}$\,cm$^{-3}$ and $2.1\times10^{11}$\,cm$^{-3}$ at 06:40:18\,UT,  06:41:30\,UT and 06:42:43\,UT, respectively.
The variation of the electron densities suggests a compressing process in the ejecta near the loop apex.

\section{Discussion and conclusions}
\label{sect:concl}
In the present work, we report on coordinated observations of a solar jet taken by SDO, IRIS and NVST.
The jet was triggered by the destabilization of a mini-filament.
The observations reveal that the ejecta of the jet are propagating along coronal loops that are as long as $\sim$70\,\arcsec.
The propagation of the ejecta away from the jet base can reach a velocity of $>$100\,\kms\ seen in the AIA 171\,\AA,
which is also confirmed by the IRIS spectral observations in \siiv\,1403\,\AA.
The ejecta can be seen in IRIS 1330\,\AA, NVST \halpha, AIA 171\,\AA\ and AIA 94\,\AA\ channels, suggesting both cool and hot components.
The emission measures of the ejecta based on the AIA EUV images show enhancements in the temperature range from log(T/K)=5.8 to log(T/K)=6.7.
Using the line pair of \oiv\ 1399.8\,\AA\ and 1401.2\,\AA, we find that the electron densities of the ejecta near the loop apex vary from $4\times10^{10}$\,cm$^{-3}$ to $2.1\times10^{11}$\,cm$^{-3}$ with a hint of occurrence of compressing processes.

\par
Before the ejecta leave the jet base region, dynamics in the remote footpoints of the coronal loops have started, including multiple compact bright kernels in \halpha\ emission and similar features in IRIS 1330\,\AA\ and AIA 171\,\AA\ passbands and a brightening in AIA 94\,\AA.
The peak of the AIA 94\,\AA\ brightenings in the remote footpoints lags 60\,s, comparing to that in the jet base.
The NoRP microwave flux at 17\,GHz and 35\,GHz also show increases in phase with the AIA 94\,\AA\ brightenings in the remote footpoints.
The occurrence of the remote brightening is followed by a sudden extension of brightness from the remote footpoint toward the jet base.
The emission measures of this bright feature based on the AIA EUV images show enhancements in the temperature range from log(T/K)=6.7 to log(T/K)=7.3.

\par
Brightenings at remote footpoints are not rare for energetic events in the corona that involve larger scale closed magnetic fields.
\citet{1992PASJ...44L.161S} reported that a short-lived X-ray bright point triggers a brightening at the remote footpoint of a loop with a seperation of 340 Mm between the two footpoints,
and that also generates a jet at the remote footpoint moving reversely toward the X-ray bright point. 
Similar phenomena appear to be common in coronal jets if they flow along closed magnetic field\,\citep{2007PASJ...59S.745S}.
Using Hinode X-ray observations, \citet{2007PASJ...59S.745S} analysed in detail one such case that generates reverse jet with a speed of 90\,\kms\ in the remote footpoint that is 480\,Mm away from the main jet.
They found a time lag of 700\,s\ between the peak time of the main jet and that of the reverse jet, and that indicates the energy is transported with a speed of 680\,\kms.
\citet{2007PASJ...59S.745S} suggested a physical mechanism for the occurrence of remote brightening and formation of reverse jet by heat conduction and/or MHD waves.
This kind of activities in remote footpoints has also been frequently seen in flares\,\citep[see e.g.][]{Wang_2012,Sun_2013,Wang_2014,Yang_2014}, and interpreted as
direct heating by nonthermal electrons at relativistic speed traveling along the large scale loops connecting the main flaring to the remote sites\,\citep[][and references therein]{2006ApJ...642.1205L}.

\par
What is the mechanism of the remote brightening seen in the present case?
The 60\,s time lag between the peak time of the jet base and that of the remote brightening indicates a speed of about 800\,\kms\ for the energy transportation, which is comparable to that found in the case reported by \citet{2007PASJ...59S.745S}.
This speed is comparable with the typical coronal Alfv\'en speed and
the propagation speed of the conduction front from flare site is also about the same speed ($\sim$500--1000\,\kms) for high temperature plasma ($\sim10^7$\,K).
The sudden extension of brightness from the remote footpoint could be a consequence of the conduction front that may excite another evaporation flow or the Alfv\'en wave that may trigger another flare (reconnection) or directly heat coronal plasma, causing counter evaporation flow. 
Alternatively, the bright kernels in the remote site in the present case have a compact round shape, which agrees with the nonthermal electron heating scenario\,\citep[see similar examples but larger scale in][]{Wang_2012}.
Evidence of nonthermal electrons produced in this coronal jet has been given by radio emissions at 17\,GHz and 35\,GHz observed by NoRP, and have also been reported in the other jets recently\,\citep{Chen_2018,Paraschiv_2019}.
The time lag between the peak times of the jet base and that of the remote footpoint is relatively large for the scenario of nonthermal electrons.
This can be interpreted as the time required to reconfigure the magnetic field in both the jet base and the remote site that allows nonthermal electrons escape and produce the brightenings\,\citep{Wang_2012}.
The nonthermal heating in the remote footpoints can then lead to chromospheric evaporation\,\citep[see e.g.][]{2015ApJ...811..139T,2016ApJ...823...41D,2016ApJ...827...27Z,2018ApJ...854...26L,2019ApJ...879L..17P,2019ApJ...879...30L}.
How such a phenomenon takes place and the role of the nonthermal electrons would be interesting subjects for detailed theoretical modelings, but that is beyond the scope of the present study.
In any case, the extension of the brightness from the remote footpoint seen in AIA 94\,\AA\ channel contains only hot plasma with temperature of about $10^7$\,K,
suggesting that they are results of the chromospheric evaporation.
The very fast propagation of the brightness could be due to the heating front that can travel much faster than the materials\,\citep{2017ApJ...849L...7D}.

\par
Based on the discussion above, we have the following scenario for the event studied here.
Initially, a mini-filament destabilization process triggers magnetic reconnection in the corona at the jet base, which is the same as that described in \,\citet{2015Natur.523..437S}.
Then, the magnetic reconnection produces high speed nonthermal electrons, MHD waves and/or conduction front that move along the large coronal loops and the jet ejecta containing materials from the mini-filaments that move much slower in the same loops.
Next, chromospheric evaporations are generated at the remote footpoints as the consequences of the interactions between the chromospheric plasma and the nonthermal phenomena produced by the magnetic reconnection processes of the jet (such as nonthermal electrons, MHD waves and/or conduction front).
After that, the chromospheric evaporations move along the coronal loops toward the jet bases and meet the jet ejecta near the loop tops.
Finally, the jet ejecta are braked by the chromospheric evaporation and reversed their moving directions.

\par
The imaging and spectral data reveal that the ejecta turn back to the jet base after crossing the apexes of the loops and meet the extension of brightness from the remote footpoints.
The resulting decelerations are found to be in the range from $1.5\pm0.8$\,km\,s$^{-2}$ to $3.0\pm1.0$\,km\,s$^{-2}$ while the ejecta are in the places with velocities approaching zero.
This indicates that in addition to gravity, there must be other forces that brake the ejecta and turn them backward.
The additional forces are very likely the pressures of the chromospheric evaporation in the remote footpoints.
%Because the extending brightness along the jet loops in AIA 94\,\AA\ is less than 10 times of the quiet regions, we can assume that its density should not exceed $10^9$\,cm$^{-3}$ (10 times of normal coronal density).
%This means that the density of the evaporation materials (i.e. the extending brightness) is a magnitude less than that of the ejecta of the jet.
If we ignore any hydrodynamics effects in the interface of the ejecta and the evaporation materials, we can have a simplified cylinder model.
With this simple model, based on the one-dimensional momentum equations for compressible fluid we can estimate the dynamic pressure ($P_{dy}^{evp}$) of the evaporated materials near the loop tops, where we can ignore the gravity, 
using the following formula:
$$P_{dy}^{evp}+ n_{evp}kT_{evp} - P_{dy}^{jet} - n_{jet}kT_{jet}=\frac{m_{jet}a}{S_{jet}}=\rho_{jet}L_{jet}a,$$
where $k$ is Boltzmann constant, $n_{evp}$ and $T_{evp}$ are the number density and temperature of the chromospheric evaporation respectively
 and the term of $n_{evp}kT_{evp}$ is the corresponding thermal pressure, 
$n_{jet}$ and $T_{jet}$ are the number density and temperature of the jet ejecta respectively
and the term of $n_{jet}kT_{jet}$ is the corresponding thermal pressure, 
$m_{jet}$ is the mass of the ejecta of the jet, $S_{jet}$ is the area of the cross-section of the ejecta,
$a$ is the deceleration of the ejecta,
$\rho_{jet}$ is the mass density of the ejecta, $L_{jet}$ is the length of the ejecta that is about 35\arcsec\ (i.e. $2.5\times10^7$\,m) in the present case,
and $P_{dy}^{jet}$ is the dynamic pressure of the jet.
The electron density of the ejecta is about $4\times10^{10}$\,cm$^{-3}$ and thus the mass density $\rho_{jet}$ is about $7\times10^{-11}$\,kg\,m$^{-3}$
if we assume the ejecta contain only electrons and protons.
$P_{dy}^{jet}$ expresses $\frac{1}{2}\rho_{jet}v_{jet}^2$ where $v_{jet}$ is the speed of the jet that is about 130\,\kms,
and thus $P_{dy}^{jet}$ is given 0.6\,Pa.
Based on the analyses of emission measures, here we use $10^6$\,K as the temperature of the ejecta and $10^7$\,K as the temperature of the chromospheric evaporation.
Because the number density of the chromospheric evaporation cannot be achieved with the present data,
we adopt the value of $10^{10}$\,cm$^{-3}$, which is the same order of that found in the chromospheric evaporations of C class flares\,\citep{2011ApJ...740...70M, 2017A&A...601A..39P}.
While $a$ is given $1.5\pm0.8$\,km\,s$^{-2}$, the dynamic pressure of the chromospheric evaporation ($P_{dy}^{evp}$) is $2.4\pm1.4$\,Pa.
While $a$ is given $3.0\pm1.0$\,km\,s$^{-2}$, the dynamic pressure of the chromospheric evaporation ($P_{dy}^{evp}$) is $5.1\pm1.8$\,Pa.
$P_{dy}$ expresses $\frac{1}{2}\rho_{evp}v_{evp}^2$, where $\rho_{evp}$ is the mass density of the chromospheric evaporation and assumed to be $1.8\times10^{-11}$\,kg\,m$^{-3}$ using the number density given above,  $v_{evp}$ is the velocity of chromospheric evaporation.
The velocity of chromospheric evaporation ($v_{evp}$), which is normally difficult to obtain due to the insufficient temporal resolution, can be deduced here via the calculations of the dynamic pressure.
For $a =1.5\pm0.8$\,km\,s$^{-2}$, $v_{evp}$ is given as $490\pm160$\,\kms.
For $a = 3.0\pm1.0$\,km\,s$^{-2}$, $v_{evp}$ is given as $750\pm130$\,\kms.
These values are consistent with those reported previously for flares based on spectroscopic data, in which the line-of-sight velocity of chromospheric evaporations are found to be a few hundred kilometers per second\,\citep[see][and the references therein]{2015ApJ...811..139T}.
Please note that the change of cross section of the flux tube and the effect of MHD waves should also play a role, but that should involve an MHD simulation to understand the physics thus could be an interesting subject for a future study.

\par
In summary, the heating in the remote footpoints by nonthermal electrons may act as a brake on the jet flows along coronal loops.
This effect prevents the transportation of plasma from the jet base to a remote site along a coronal loop.
The changes in the motions of the ejecta caused by the brake can then allow a diagnostics to the dynamics of the plasma heated in the remote footpoints. 

%\acknowledgments
{\it Acknowledgments:}
We are grateful to the referee for the constructive suggestions and comments that help improve the manuscript. We also thank Dr. Shiwei Feng and Dr. Rongsheng Wang for useful discussions.
This research is supported by National Natural Science Foundation of China (U1831112, 41974201, 41604147, 11761141002),
and the Young Scholar Program of Shandong University, Weihai (2017WHWLJH07).
The NVST data can be freely requested from the official website at \textit{http://fso.ynao.ac.cn}.
We would like to thank the NVST operation team for preparation of the data.
IRIS is a NASA small explorer mission developed and operated by LMSAL with mission operations executed at NASA Ames Research center and major contributions to downlink communications funded by ESA and the Norwegian Space Centre.
Courtesy of NASA/SDO, the AIA and HMI teams and JSOC.
CHIANTI is a collaborative project involving George Mason University, the University of Michigan (USA) and the University of Cambridge (UK).

\bibliographystyle{aasjournal}
\bibliography{bibliography}

\end{document}